# Parametric reduced order models with machine learning for spatial emulation of mixing and combustion problems

Chenxu Ni, Siyu Ding, Xingjian Wang[*]

Department of Energy and Power Engineering, Tsinghua University, Beijing 100084,

People's Republic of China

**Abstract**

High-fidelity simulations of mixing and combustion processes are generally computationally demanding and time-consuming, hindering their wide application in industrial design and optimization. The present study proposes parametric reduced order models (ROMs) to emulate spatial distributions of physical fields for multi-species mixing and combustion problems in a fast and accurate manner. The model integrates recent advances in experimental design, high-dimensional data assimilation, proper-orthogonal-decomposition (POD)-based model reduction, and machine learning (ML). The ML methods of concern include Gaussian process kriging, second-order polynomial regression, k-nearest neighbors, deep neural network (DNN), and support vector regression. Parametric ROMs with different ML methods are carefully examined through the emulation of mixing and combustion of steam-diluted fuel blend and oxygen issuing from a triple-coaxial nozzle. Two design parameters, fuel blending ratio (hydrogen/methane) and steam dilution ratio, are considered. Numerical simulations are performed and training data is assimilated at sampling points determined by the Latin-hypercube design method. The results show that ROM with kriging presents a superior performance in predicting almost all physical fields, such as temperature, velocity magnitude, and combustion products, at different validation points. The accuracy of ROM with DNN is not encouraging owing to the stringent requirement on the size of training database, which cannot be guaranteed as many engineering problems are specific and associated data availability is limited. For the emulation of spatial field, the parametric ROMs achieve a turnaround time of up to eight orders of magnitude faster than conventional numerical simulation, facilitating an efficient framework for design and optimization.

---

[*] Corresponding author.
E-mail address: xingjianwang@tsinghua.edu.cn

**Novelty and significance statement**

The present study proposes parametric reduced order models (ROMs) to emulate spatial distributions of physical fields for multi-species mixing and combustion problems in a fast and accurate manner. The ROMs incorporate projection-based dimension reduction, machine learning (ML)-based regression techniques, and Latin-hypercube sampling design method into a unified framework. The developed ROMs are data-driven and expected to provide reliable surrogates of numerical simulations in a much shorter turnaround time and a more cost-effective way, achieving speeds up to eight orders of magnitude faster than conventional numerical simulation. The results show that ROM with kriging presents a superior performance in predicting almost all physical fields, and can also be applied to a wide range of engineering applications.

**Authors contributions**

**Chenxu Ni:** Methodology, Software, Visualization, Writing-original draft. **Siyu Ding:** Software, Writing-

review & editing. **Xingjian Wang:** Conceptualization, Supervision, Writing – review & editing.

## 1. Introduction

This paper attempts to develop and compare multiple parametric reduced-order models (ROMs) for emulating the spatial field of fuel-flexible oxy-combustion processes at different steam-dilution ratios and hydrogen blending ratios. The ROMs incorporate projection-based dimension reduction, machine learning (ML)-based regression techniques, and Latin-hypercube sampling design method into a unified framework. The developed ROMs are data-driven and expected to provide reliable surrogates of numerical simulations in a much shorter turnaround time and a more cost-effective way. This is useful for the design optimization of hydrogen-rich gas-turbine (GT) combustors, where numerous design parameters are required to survey in a wide design space.

Advances in optical diagnostics, image processing, and high-performance computing have enabled us to gain an increasingly better understanding of combustion processes and flame dynamics in practical problems. Experimental studies often come with great costs and high risks, while numerical investigations are computationally expensive and time-consuming. Through these physical experiments and computer simulations in the past many years, an enormous amount of data has been accumulated. Primary sources of experimental data include fundamental thermochemical properties, spatiotemporal images to visualize flame evolutions, and sensing information for control and health monitoring of energy-conversion systems. In regard to simulation data, Ihme et al. [1] conducted a comprehensive review of direct numerical simulation (DNS) studies on turbulent reacting flows in the past two decades. The size of cumulative data to describe flame solution fields is on the order of 10 petabytes. The gigantic data pool offers great opportunities for multidisciplinary research, especially with the help of big data and artificial intelligence.

In the design optimization of many-query engineering systems, the survey of a large parametric space and multiple design attributes is required. For example, globally stringent emission regulation on carbon dioxide and NOx has shifted conventionally hydrocarbon-fueled GT combustors towards pure hydrogen or hydrogen-rich technology. The latter is considered as one of the promising choices for future zero-carbon society to regulate peak demand and mitigate the intermittency of renewable power grid system [2]. However, hydrogen GT exhibits many new operability challenges in the context of present lean-premixed technology, including flashback, autoignition, and combustion stability. The unique characteristics of hydrogen flames require to be fully concerned in the design process. This may involve quite a few design parameters, such as geometric parameters, fuel-oxidizer equivalence ratio, injection strategy, cooling system, and so forth. As a result, relying solely on high-fidelity simulations and/or experiments for this task becomes impractical. To this end, data-enabled surrogate models have been developed to accurately emulate the results of numerical simulations in a significantly reduced time window, thus improving design efficiency.

Early surrogate models were primarily data-fit type, which are not physics-based and are formulated directly by constructing a response surface using data at a few design points [3]. Traditional response models, such as kriging, radial basis functions, support vector regression (SVR), and multivariate polynomial, have been widely used in design analysis and optimization [4-6]. These models, however, face the challenge of the "curse of dimensionality" issue in the case of high-dimensional problems. Artificial neural network (ANN)-based data-fit models have recently demonstrated good capability for high-dimensional and large gradient tasks [7-9]. For multi-physics and multiscale engineering problems with parametric variability, however, data availability is often limited (both computationally and experimentally). The ANN-based data-fit models with small dataset may deviate from the underlying physical principles governed by partial differential equations (PDEs), and the resultant prediction performance may be compromised.

Another type of surrogate model is reduced-order model (ROM), which creates a low-dimensional subspace consisting of a set of basis vectors determined from the original high-dimensional data. Various ROMs have been proposed. Autoencoders, as a type of ANN structure, have been implemented to develop ROMs. An autoencoder consists of an encoder network that maps the high-dimensional input to a lower-dimensional manifolds, and a decoder network that reconstructs the high-dimensional input from the encoded representation [10-14]. The curse of dimensionality issue is overcome by the encoder procedure, where the latent variables in a significantly reduced dimension are deduced from the observational data (input). Milano and Koumoutsakos [15] demonstrated that autoencoders offer enhanced reconstruction and prediction capabilities for near-wall velocity distributions, at a marginal increase in computational cost compared to POD. While autoencoders have the ability to capture nonlinear trial manifolds through complex network structures and activation functions, their correspondence to actual dynamical information is not well understood. Projection-based ROMs, on the other hand, currently the most widely used in engineering, provide a direct link between the low-dimensional subspace and underlying flow dynamics, thereby retaining physical interpretability.

The low-dimensional subspace in projection-based ROMs can be approached by proper orthogonal decomposition (POD), also known as principal component analysis [16] in the scientific computing community. POD divides data into a set of orthonormal basis functions (known as POD modes) and associated time-varying coefficients [17]. The POD modes span a low-dimensional subspace and represent the most dominant flow dynamics. Various regression techniques are then employed to learn the mapping from parametric inputs to modal coefficients at new design points. This method provides a systematic way for effective and accurate reconstruction of multi-physics and multiscale system, while avoiding potential deviations from physics that may arise in purely data-fit models. Extensive studies have been performed on developing the POD-based ROMs in various research topics, including the continuous efforts made by Yang and his colleagues [18-22] for spatiotemporal problems, moving domain problems [23-25], aerodynamics [26-28], and combustion [29-36].

However, most efforts made so far mainly concentrated on the development of POD modes (low-dimensional subspace) and drew little attention to the accuracy evaluation of the regression technique to establish the mapping between the parametric input and ROM output. For the present mixing and reacting problems, multiple outputs, including combustion products, temperature, and velocity magnitude, exhibit different features for a given design point. The selection of regression methods to define the mapping function and the optimal number of POD modes spanning the reduced subspace require to be carefully calibrated. The impact of different ML methods on the ROM predictor could be problematic when the input-output map is of intermediate or high gradient (due to the need for access to the high-dimensional operators). For instance, for a single-injector combustion problem, it requires more than 100 modes to obtain stable ROMs with sufficient accuracy [30]. Besides, the prediction of minor species is challenging owing to their relatively small magnitude in the products. This necessitates a comprehensive investigation of the regression technique to determine the mapping between the parametric input and the modal coefficients. The objective of the present study is to develop parametric POD-based ROMs capable of physical interpretability and suitable for flows and combustion problems using ML approaches.

The paper is organized as follows: Section 2 introduces the triple-coaxial injector configuration, the numerical framework, and sample points determined by design of experiment (DoE). Section 3 details the framework of the POD-based ROM and introduces the ML methods coupled to the POD algorithm. Section 4 discusses the results for the two model problems, namely, the mixing and combustion of fuel blend (methane/hydrogen), stream, and oxygen triple-coaxial jets, the accuracy and computational efficiency of different ML methods are discussed and analyzed in comparison to baseline numerical simulations. Section 5 concludes this work.

## 2. Physical model description and data collection

### 2.1 Triple-Coaxial Injector Configuration

In this study, we demonstrate the parametric ROMs based on the modeling of a single micro-mixed nozzle configured in a multi-feed jet injection pattern, as shown in Fig. 1. Fuel blend (a mixture of hydrogen and methane) is injected in the inner circular jet, while steam and oxygen are coaxially injected in the middle and outer circular jets in the radially outward direction, respectively. Such triple-coaxial (for brevity, tri-coaxial hereafter) injection issuing from a nozzle can mitigate the flame temperature conveniently by adjusting the stream/oxygen mass flow rates. Additionally, a co-flow steam is injected surrounding the tri-coaxial nozzle. The wall thickness between the adjacent circular jets is 1 mm. The geometric conditions of the tri-coaxial nozzle are tabulated in Table 1, where $D_i$ and $D_o$ denote inner and outer diameters of each concentric jet, respectively.

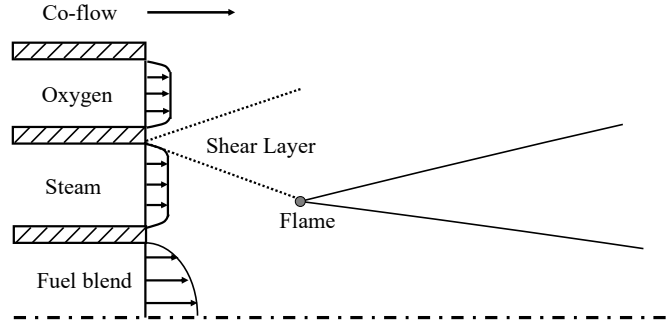

Fig. 1 Schematic diagram of a triple-coaxial nozzle

Table 1 The nozzle geometric conditions

|  | $D_i(mm)$ | $D_o(mm)$ |
|---|---|---|
| Fuel | N/A | 9 |
| Steam | 11 | 16 |
| Oxygen | 18 | 22 |
| Co-flow | 24 | 160 |

## 2.2 Design of experiment

Design optimization process of a multi-query problem can be facilitated by design of experiment (DoE) [37]. DoE is a powerful statistical tool to determine the sampling distribution of input variables for a given space. The present study considers two design variables $x = \{X_{CH_4}, \emptyset_{H_2O}\}^T \in \mathbb{R}^2$: the mole fraction of methane in the range of 0-1, $X_{CH_4}$ and the volume fraction of water vapor in the range of 0-0.5, $\emptyset_{H_2O}$, which is defined as the ratio of the steam volumetric flow rate and the total volumetric flow rate of the tri-coaxial nozzle, namely:

$$\emptyset_{H_2O} = \frac{q_{\langle H_2O \rangle}}{q_{\langle fuel \rangle} + q_{\langle H_2O \rangle} + q_{\langle O_2 \rangle}} \quad (1)$$

where $q_{\langle fuel \rangle}$, $q_{\langle H_2O \rangle}$, and $q_{\langle O_2 \rangle}$ represent volumetric flow rates of fuel, steam, and oxygen issuing from the tri-coaxial nozzle, respectively.

The sample size is determined based on the 10*d rule-of-thumb* [38]. Here, *d* represents the number of design variables. Once the design variables and the sample size are selected, we employ the *Latin hypercube sampling* (*LHS*) method to generate the design points. Fig. 2 shows the design points calculated by the *LHS*. The sampling design points are well fulfilled in the design space, which is crucial in constructing ROM. When setting the inlet boundary conditions for numerical simulations, there are two constraints to be satisfied based on the design variables: the equivalence ratio of fuel and oxygen should be close to the stoichiometric ratio, and the inlet velocity of fuel remains unchanged. Table 2 lists the design points and corresponding inlet boundary conditions (inlet velocity of oxygen, $v_{O_2}$ and inlet velocity of steam, $v_{H_2O}$). The last two design points are used as the validation set, namely test cases 1 ($X_{CH_4} = 0.3, \emptyset_{H_2O} = 0.15$) and 2 ($X_{CH_4} = 0.5, \emptyset_{H_2O} = 0.25$).

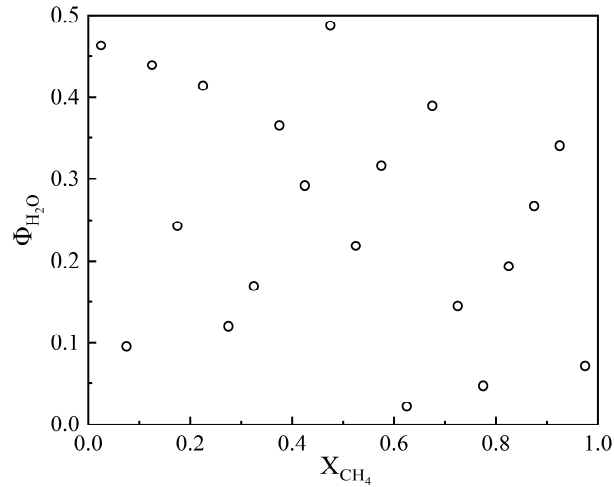

Fig. 2 Distribution of design points obtained by *LHS* in the design space

Table 2 The design points and corresponding inlet boundary conditions

|    | $X_{CH_4}$ | $\emptyset_{H_2O}$ | $v_{O_2}$ | $v_{H_2O}$ |    | $X_{CH_4}$ | $\emptyset_{H_2O}$ | $v_{O_2}$ | $v_{H_2O}$ |
|----|------------|--------------------|-----------|------------|----|------------|--------------------|-----------|------------|
| 1  | 0.375      | 0.3653             | 22.95     | 30.38      | 12 | 0.725      | 0.1448             | 34.29     | 11.21      |
| 2  | 0.325      | 0.1693             | 21.33     | 10.37      | 13 | 0.675      | 0.3897             | 32.67     | 41.06      |
| 3  | 0.875      | 0.2672             | 39.15     | 26.25      | 14 | 0.075      | 0.0957             | 13.23     | 4.37       |
| 4  | 0.925      | 0.3407             | 40.77     | 38.19      | 15 | 0.025      | 0.4632             | 11.61     | 33.96      |
| 5  | 0.125      | 0.4387             | 14.85     | 33.76      | 16 | 0.525      | 0.2182             | 27.81     | 16.34      |
| 6  | 0.475      | 0.4878             | 26.19     | 53.93      | 17 | 0.975      | 0.0712             | 42.39     | 5.81       |
| 7  | 0.425      | 0.2918             | 24.57     | 22.54      | 18 | 0.275      | 0.1202             | 19.71     | 6.69       |
| 8  | 0.625      | 0.0222             | 31.05     | 1.42       | 19 | 0.575      | 0.3163             | 29.43     | 27.98      |
| 9  | 0.775      | 0.0467             | 35.91     | 3.34       | 20 | 0.175      | 0.2427             | 16.47     | 14.46      |
| 10 | 0.825      | 0.1938             | 37.53     | 16.84      | 21 | 0.300      | 0.1500             | 20.52     | 8.81       |
| 11 | 0.225      | 0.4143             | 18.09     | 33.27      | 22 | 0.500      | 0.2500             | 27.00     | 19.20      |

## 2.3 Numerical simulations

To train a POD-based ROM in a non-intrusive manner, a pre-calculated database containing the simulation results at the design points sweeping over the sampling design space is required. Computations were performed using the Reynolds-averaged Navier-Stokes (RANS) approach. Turbulence closure is achieved using the $SST\ k-\omega$ model [39]. In combustion cases, the closure of chemical reaction rates is modeled by the finite-rate/eddy-dissipation model. Eddy-dissipation model assumes reaction rates to be controlled by the turbulent mixing and ignores chemical time scales [40] This is usually acceptable for non-premixed flames, but for the present configuration in Fig. 1, a partially premixed region near the nozzle exit exists to stabilize the flame. The finite-rate model computes the reaction rates according to the Arrhenius law and acts as a kinetic switch. The GRI 3.0 mechanism [41] is implemented for chemistry modeling, consisting of 53 species and a total of 325 reversible reactions. Note that more advanced turbulence/combustion modeling techniques may be adopted, this is, however, not the focal point of the present study. The purpose of numerical simulations is to provide a comprehensive database for the following parametric ROM development. The algorithm of ROM itself can be applied to any dataset available.

The numerical framework employs a pressure-based, finite-volume methodology, along with a semi-implicit method for pressure linked equations (SIMPLE) scheme. Spatial discretization is based on a second-order upwind scheme. The inlet temperature of fuel and oxygen is fixed at 300 K, while the steam inlet is set at 500 K. Besides, the fuel inlet and the co-flow velocity remain fixed at 42.66 $m/s$ and 0.3 $m/s$, respectively. Other inlet boundary conditions are based on the design variables calculated from the DoE.

The computational domain includes the nozzle and a downstream region spanning 360 mm in the axial direction and 160 mm in the radial direction. Axisymmetric calculations are performed due to the high computational cost when considering the detailed chemical kinetics of hydrogen/methane combustion. To address the rapid flow variations in the shear layers and near the nozzle exit, more grid points are clustered in these regions. The resultant number of computational grids is 93,600.

Numerical simulations of non-reacting and reacting cases were performed over all training design points and testing points. As a specific example, Fig 3 illustrates the temperature distribution in reacting cases at three different design points. As the volumetric flow rate of steam increases, the lift-off distance increases. Increased steam dilution slows the chemical reaction rates near the nozzle exit and thus shifts the flame anchoring point further downstream. Another interesting observation for the OH distribution is that the flame length becomes shorter with decreasing methane mole fraction (and increasing hydrogen mole fraction) in the fuel blend. This is one of the salient features of hydrogen-rich flames, compared to methane flames

[42].

To construct the database for the parametric ROM, the spatial domain of interest is tailored to cover the axial direction from 20 to 360 mm and the radial direction from the centerline to 60mm. For the non-reacting cases, the objective of the ROM is to emulate the distributions of velocity magnitude (*v*), the mole fractions of primary species (such as $H_2$, $CH_4$, and $O_2$). For the reacting cases, the variables of interest selected for the emulation include temperature (*T*), *v*, and important species mole fractions (such as $H_2, O_2, H_2O, CO_2, CH_2O$ and $OH$).

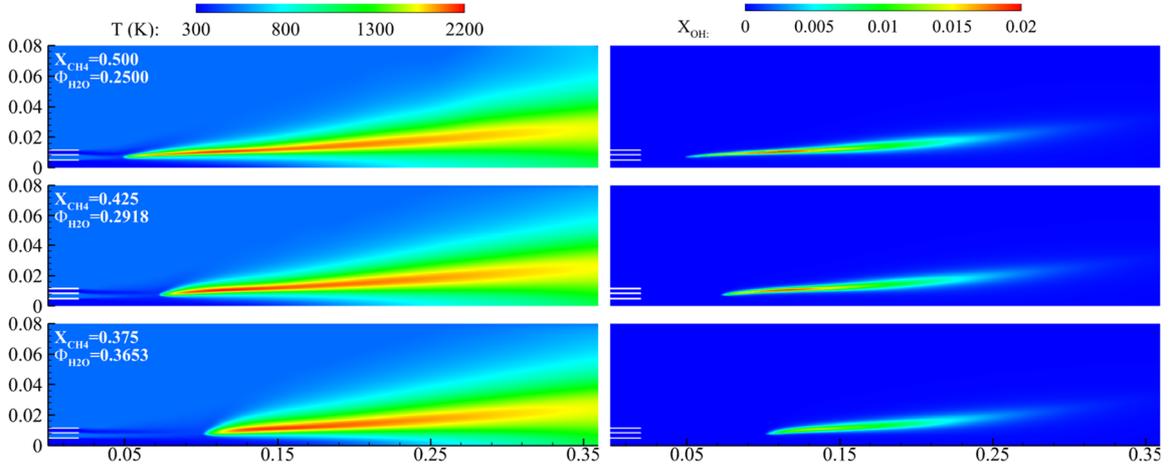

Fig. 3 Temperature field of combustion cases at different design points.

## 3. POD-based parametric ROM

Having the collected data above, this section describes the theory of the POD-based parametric ROM. Figure 4 shows the methodology of parametric ROM for the present study. POD is implemented to extract the dominant flow features (POD modes) embedded in the training dataset. Mathematically, this operation produces a low-dimensional subspace, where the dataset in all training cases is projected. The coefficients associated with the modes can be interpreted as the coordinates in the subspace. The core of the parametric ROM is to configure the coordinates of unknown design points based on those of the training points. To this end, we propose several ML techniques to optimize the coefficients and evaluate the performance of various ROMs. In the following, the POD method is first briefed, followed by the introduction of various ML methods applied in this work.

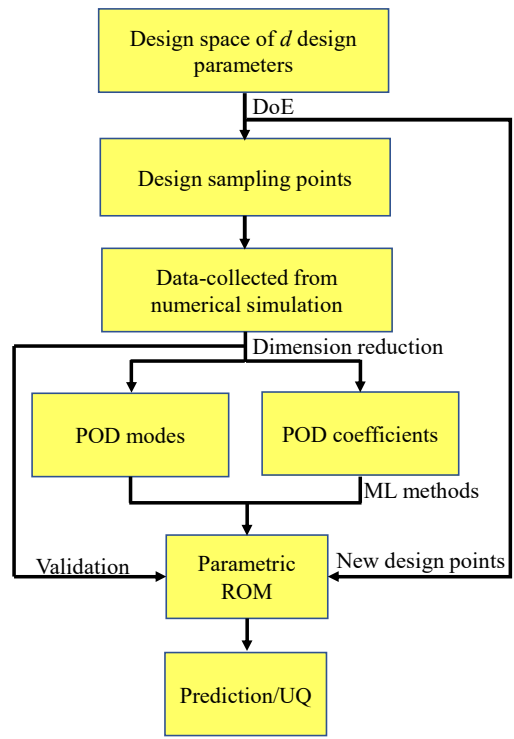

Fig. 4 POD-based parametric ROM methodology

**3.1 POD**

The POD method is widely used in modal analysis of flow and combustion dynamics. For a given flowfield dataset, POD identifies a set of orthonormal basis functions (POD modes), where the least-square error of the dataset projection is minimized. The selection of these basis functions is based on the magnitude of the associated eigenvalues. The POD modes with large eigenvalues denote significant dynamic structures in the data. It is natural to configure a subspace including the POD modes with large eigenvalues while cutting off those with low eigenvalues, and this reduced subspace is supposed to recover the most important information in the original dataset with minimal uncertainty.

To elaborate on this, a snapshot-based approach is employed [43]. We assume the training dataset can be represented by a snapshot matrix $Y = \{y^{(i)}, i = 1, 2, ..., n\} \in \mathbb{R}^{m \times n}$, where $y^{(i)} \in \mathbb{R}^m$ is the data information at design point $x^{(i)} \in \mathbb{R}^d$, $m$ the number of selected grid points at design point $i$, and $n$ the total number of design points (snapshots). $m$ is normally much larger than $n$, rendering the matrix $Y$ a tall and thin shape. Gram matrices $YY^T$ ($\in \mathbb{R}^{m \times m}$) and $Y^T Y$ ($\in \mathbb{R}^{n \times n}$) are inner products of row vectors and column vectors of the matrix $Y$, respectively. Both Gram matrices can be used to find eigenvalues and eigenvectors (basis functions) through eigendecomposition. To avoid high dimensionality, $Y^T Y$ is selected here. The snapshot-based method solves an eigenvalue problem as follows:

$$Y^T Y \psi_j = \lambda_j \psi_j, \qquad \psi_j \in \mathbb{R}^n \tag{2}$$

The eigenvalues $\lambda_j$ guide the determination of the number of the POD modes required for an accurate representation of the original dataset, and they are sorted in a descending order. The modes with the largest eigenvalues capture the most significant variability in the dataset. The dimension reduction is achieved by selecting the first $r$ ($r < n$) eigenvalues and cutting off the reminder. A common approach is to determine $r$ as follows:

$$\frac{\sum_{j=1}^{r} \lambda_j}{\sum_{j=1}^{n} \lambda_j} > \delta \tag{3}$$

where $\delta$ is a user-defined tolerance, taken to be 99% in our study. The eigenvectors $\psi_j \in \mathbb{R}^n$ can deduce the eigenvectors of the Gram matrix $YY^T$, that is, $\phi_j \in \mathbb{R}^m$ as follows:

$$\phi_j = Y \psi_j \frac{1}{\sqrt{\lambda_j}} \in \mathbb{R}^m \tag{4}$$

The POD mode matrix $\Phi$ can be configured by an array of $\phi_j$, $\Phi = \{\phi_j, j =$

$1, 2, \ldots, r\} \in \mathbb{R}^{m \times r}$. The mode matrix is a reduced subspace ($m \times r$) spanning $r$ column vectors ($\boldsymbol{\phi}_j$). The projection of the training data at design point $i$ into this subspace gives the coefficient vector (or coordinates $\boldsymbol{\beta}^{(i)} = \left(\beta_1^{(i)}, \beta_2^{(i)}, \ldots, \beta_r^{(i)}\right)^T \in \mathbb{R}^r$ associated with the data in the subspace. It is thus natural to recover the information in the training dataset using the following way:

$$y^{(i)} = \sum_{j=1}^{r} \beta_j^{(i)} \boldsymbol{\phi}_j = \boldsymbol{\Phi} \boldsymbol{\beta}^{(i)} \tag{5}$$

Equation (5) can be thought of as a ROM of the original data in the training set. To develop parametric predictability, the coefficient vector ($\boldsymbol{\beta}^{(new)}$) at unknown design points ($\boldsymbol{x}^{(new)}$) is required to find out based on existing information in the training samples. The following subsection introduces several ML methods to establish a predictive model for the POD coefficients, which enables the estimation of spatial fields at new design points.

### 3.2 Machine learning

ML methods can be used to predict the modal coefficient vector ($\boldsymbol{\beta}^{(new)}$)) due to the ability to capture complex nonlinear relationships between input parameter and output response, to adaptively learn from the available training samples, and to generalize well to unseen data. We explore the applicability of five different ML techniques to the present mixing and combustion problem, including kriging, k-nearest-neighbor (KNN) regression, multivariate polynomial regression (MPR), deep neural network (DNN), and support vector regression (SVR). These techniques are briefly discussed in the following.

#### 3.2.1 Kriging

Kriging, also known as Gaussian process (GP) regression, is a statistical interpolation technique used for spatial modeling and prediction [44]. The basic principle of kriging involves modeling the spatial variability between sample data points using a covariance function, fitting the covariance parameters using the sample data, and then using the covariance to calculate the weights of the sample data to estimate the value at the target point. Therefore, kriging can accurately approximate the response surface over the entire design space using output values sampled at a set of input parameters. By carefully choosing the model parameters, kriging can provide the best linear unbiased estimator for responses at unknown design points, even for unsampled regions of the response surface.

Kriging treats deterministic output responses as realizations of a random process. For the present problem, the coefficient vector ($\boldsymbol{\beta}^{(i)}$) at training sample $\boldsymbol{x}^{(i)}$ is randomized with a multivariate normal distribution,

$$\boldsymbol{\beta}^{(i)} \sim GP(\boldsymbol{\mu}; \boldsymbol{\Sigma}) \tag{6}$$

where $\boldsymbol{\mu} \in \mathbb{R}^r$ denotes the process mean vector of the response $\boldsymbol{\beta}^{(i)}$ at sampled design points, and $\boldsymbol{\Sigma}: \mathbb{R}^d \times \mathbb{R}^d \to \mathbb{R}^{r \times r}$ the corresponding covariance matrix function, a squared-exponential form:

$$\boldsymbol{\Sigma}(\boldsymbol{x}^{(i)}, \boldsymbol{x}^{(j)}) = \exp\left(-\sum_{k=1}^{d} \theta_k \left(x_k^{(i)} - x_k^{(j)}\right)^2\right), \tag{7}$$

where $\theta_k$ quantifies the level of correlation among the data in the $k^{\text{th}}$ component of design parameters. Since the modal coefficient vectors $\boldsymbol{\beta}^{(i)}, i = 1, 2, \ldots, r$ at $\boldsymbol{x}^{(i)}$ are known, invoking the conditional distribution of the multivariate normal distribution gives the predictor of the coefficient vector $\widehat{\boldsymbol{\beta}}^{(new)}$ at a new design point, which follows the form:

$$\widehat{\boldsymbol{\beta}}^{(new)} = \boldsymbol{\mu} + \boldsymbol{c}^T \boldsymbol{C}^{-1} \otimes \boldsymbol{I}_r (\boldsymbol{\beta} - \boldsymbol{1}_n \otimes \boldsymbol{\mu}), \tag{8}$$

where the coefficient vector $\boldsymbol{\beta} = \left( \left(\boldsymbol{\beta}^{(1)}\right)^T, \left(\boldsymbol{\beta}^{(2)}\right)^T, \ldots, \left(\boldsymbol{\beta}^{(n)}\right)^T \right)^T \in \mathbb{R}^{nr}$ consists of the coefficient vectors $\boldsymbol{\beta}^{(i)}$ of the training samples, and $c$ is an n-row vector identifying the correlation between the new design point and the training sampling points. $C$ is a $n \times n$ correlation matrix of the training points, and $\mathbf{1}_n$ is an n-row vector of ones.

### 3.2.2. k-nearest-neighbor regression (KNN)

KNN is a simple and widely used supervised learning technique. The basic idea of the KNN algorithm is to find the $k$ nearest neighbors of a new design point in the design space, based on a distance metric such as Euclidean distance. The coefficient vector ($\boldsymbol{\beta}^{(new)}$) at new design points ($\boldsymbol{x}^{(new)}$) is then determined by the averaged values of its $k$ nearest neighbors.

The hyperparameter $k$ is selected through the cross-validation process, which divides the dataset into multiple folds and optimizes $k$ on different combinations of training and validation sets. The model is trained on a subset of folds and evaluated on the remaining fold with various values of $k$. This process is repeated multiple times, with each fold being used as a validation set once. The performance of the model, using the root mean square error $E$ as the metric, is averaged over all the validated folds to obtain a more robust estimate. The final value of $k$ is determined to give the best performance metric. The steps for selecting the value of $k$ in KNN using cross-validation are illustrated in Fig. 5.

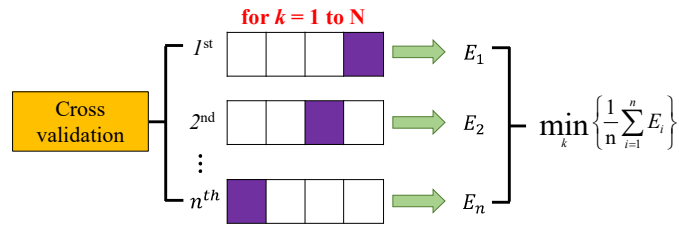

Fig. 5 Cross-validation methodology of KNN regression

### 3.2.3. Multivariate polynomial regression (MPR)

Multivariate polynomial regression extends the concept of polynomial regression to multiple independent variables, an attempt to capture complex nonlinear relationships between variables. It establishes the relationship between the parametric input $x^{(i)}$ and the mode coefficients $\beta^{(i)}$ as a polynomial equation based on the least squares regression. In this paper, the quadratic formulation of our model is written as:

$$\beta^{(new)} = a_0 + \sum_{i=1}^{d} b_i x_i^{(new)} + \sum_{i=1}^{d}\sum_{j=i}^{d} c_{ij} x_i^{(new)} x_j^{(new)} \tag{9}$$

where $a_0$, $b_i$, $c_{ij} \in \mathbb{R}^r$ are the coefficient vectors of the quadratic model determined via least squares regression.

### 3.2.4. Deep neural network (DNN)

DNN is widely recognized for its capability to effectively approximate intricate and nonlinear processes. In the DNN model, a mapping function $\beta = f(x; \theta)$ is established, and the model parameters $\theta$ consisting of the weights and the bias $\{W^{(i)}, b^{(i)}, i = 1,2,\cdots l\}$, are iteratively optimized using a training database and a suitable loss function. Thus, we can define the intermediate information carried by the neurons in the $i^{th}$ layer $h^{(i)}$ as follows:

$$h^{(i)} = f^{(i)}(h^{(i-1)}; \theta^i) = \sigma_i(W^{(i)} h^{(i-1)} + b^{(i)}) \tag{10}$$

Here, $\sigma_i$ denotes the activation function of the $i^{th}$ layer to introduce nonlinearity, and ReLU is selected in this study.

Given a training set $N_{train}$, the loss function $J(\theta)$ is taken as the mean-squared error (MSE) loss:

$$J(\theta) = \frac{1}{N_{train}} \sum_{i \in N_{train}} (\beta^{(i)} - \widehat{\beta}^{(i)})^2 \tag{11}$$

The loss function represents the degree of deviation between the actual data and the model estimations. To minimize the loss function, the Adam optimization algorithm [45] based on adaptive estimates of lower-order moments is used to update $\boldsymbol{\theta}$. This iterative process loops until the error threshold or the set number of iterations is reached.

**3.2.5 Support vector regression (SVR)**

Different from the traditional regression models, SVR calculates the estimation loss using the $\epsilon$-insensitive loss function. In SVR, the goal is to minimize the prediction error within a tolerance band ($\epsilon$) around the actual target values. Data points that fall within the tolerance band do not contribute to the loss, so the $\epsilon$-insensitive loss function $\mathcal{L}_\epsilon$ can be written as follows:

$$\mathcal{L}_\epsilon = \begin{cases} 0, & if\ |z| \leq \epsilon; \\ |z| - \epsilon, otherwise \end{cases} \quad (12)$$

The optimization objective can be formalized as:

$$\min_{\boldsymbol{\omega},b} \frac{1}{2}\|\boldsymbol{\omega}\|^2 + C \sum_{i=1}^{n} \mathcal{L}_\epsilon\big(f(\boldsymbol{x}^{(i)}) - \boldsymbol{\beta}^{(i)}\big),$$
$$f(\boldsymbol{x}^{(i)}) = \boldsymbol{\omega}^T \alpha\big(\boldsymbol{x}^{(i)}\big) + b, \quad (13)$$

where $\boldsymbol{\omega}$ is the vector normal to the direction of the hyperplane, $b$ is the bias term describing the distance between the hyperplane and the origin, $C$ represents the regularization parameter, and $\alpha(\boldsymbol{x}^{(i)})$ represents the kernel function. Here, we employed the Gaussian kernel function as the mapping function.

During the model training, the Lagrange multipliers are introduced to formulate the problem as a constrained optimization with the Lagrange function. This is followed by solving the dual problem using the Karush-Kuhn-Tucker conditions to obtain the optimal Lagrange multipliers for the solution of SVR.

## 4. Results and discussion

### 4.1. POD modes and sensitivity analysis

The POD-based ROM is known for its ability to incorporate physical information of the flowfield into the model. Figure 6 illustrates the first four POD modes of the temperature field and associated energy percentages in the combustion case. The first mode, which contributes 52% of the total energy, plays a significant role in regulating the temperature distribution of the flame zone and centerline flow recirculation in the downstream region. Mode 2 with roughly 31% in energy percentage is likely related to the variation of flame anchoring point among different design points. Modes 3 and 4 exhibit temperature fluctuations primarily concentrated in the mixing layer near the multi-feed nozzle exit where shear-layer vortices develop. Note that a POD mode may not always represent an actual flow coherent structure. A flow pattern can only exhibit prominence in a mode when the associated physical event remains active across all training datasets.

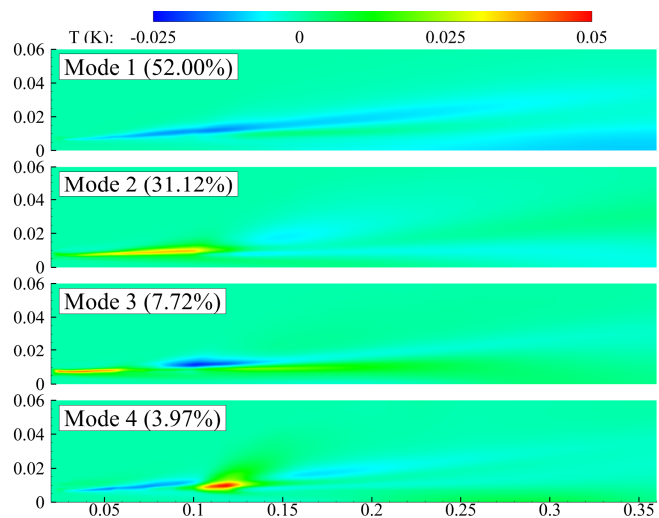

Fig. 6 Contour plots of first four POD modes of temperature field in combustion case.

The number of selected POD modes is critical to the prediction accuracy of the parametric ROM. This selection can be measured by the energy percentage occupied by these POD modes.

The energy percentage can be expressed as the ratio of the diagonal element, corresponding to the specific POD mode, to the trace of the diagonal eigenvalue matrix of $Y^T Y$. Figure 7 shows the individual and accumulated energy percentages of POD modes for the hydrogen mole fraction ($X_{H_2}$) field for both mixing and combustion cases. Compared to the mixing case where the first two modes occupy more than 99% of the total energy, the energy accumulation is slightly slow in the combustion case (the first five modes for 99% of the total energy). This is attributed to the inherent complexity of combustion problems, which encompass multi-physics and multi-scale effects, thereby requiring a greater number of POD modes to capture the main characteristics of the physical field.

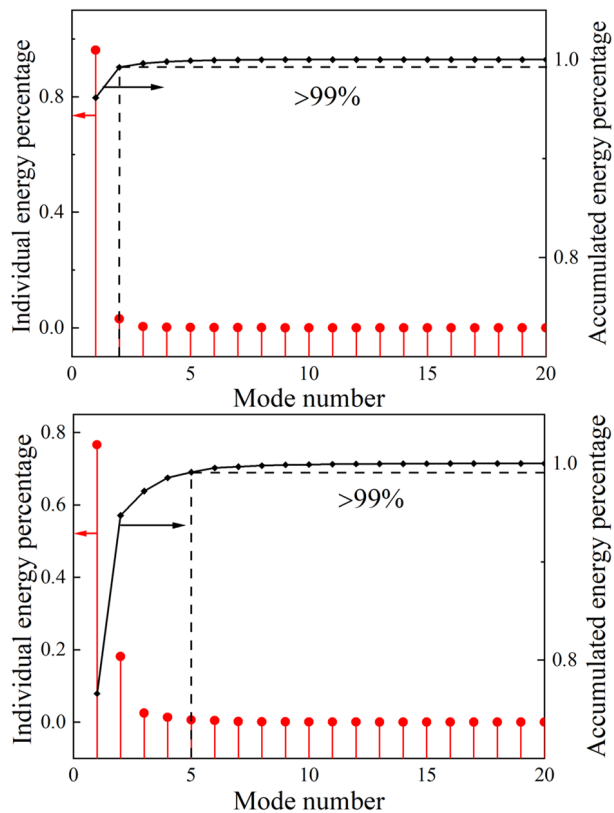

Fig. 7 Individual and accumulated energy percentage of POD modes for the $X_{H_2}$ field in the mixing (upper) and combustion (lower) problems.

In certain scenarios, building ROM solely relying on energy percentage may not be sufficient enough. In addition to the criterion of the accumulated energy percentage, a sensitivity analysis is carried out to examine the number of modes on the performance of ROM. We consider the first *r* modes, which cover about 99% percentage of the total energy, as the baseline. Subsequently, the number of modes is gradually increased, and the accuracy of the predictions is evaluated using the absolute average relative deviation (AARD).

$$AARD = \frac{1}{m}\sum_{j=1}^{m}\frac{|y_{p,j} - y_{p,j}^*|}{|y_{p,j}|} \tag{14}$$

where *p* and *j* stand for the physical field index and the number of data points, respectively. $y_{p,j}$ and $y_{p,j}^*$ are the results of RANS and the ROM prediction, respectively. For demonstration, kriging is used for sensitivity analysis.

Figure 8 illustrates the number of POD modes obtained from the baseline and sensitivity analysis. For some flow variables, the baseline according to the energy percentage of 99% is as accurate as the result of sensitivity analysis. For many other variables, such as oxygen and hydrogen mole fractions in the mixing case and temperature and hydrogen mole fraction in combustion cases, more modes are required to be incorporated into the ROM to ensure a smaller AARD. Following this way, the number of POD modes to construct the ROM is further optimized.

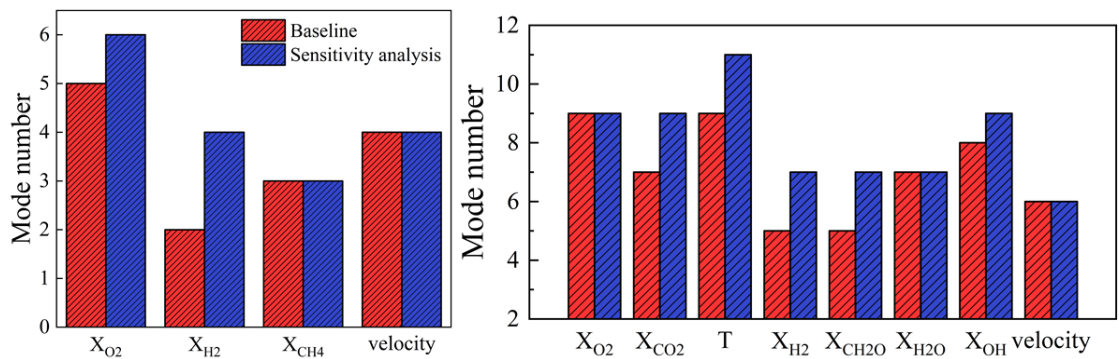

Fig. 8 Number of POD modes based on baseline and obtained from sensitivity analysis in mixing (left) and combustion (right) cases

### 4.2. ROM of mixing case

For surrogate models, it is essential to quantitatively evaluate the uncertainty associated with predictions to assess model accuracy. In this study, we assume that the database generated by RANS is reliable and accurate. Hence, the uncertainty primarily arises from the application of various ML techniques. Two performance metrics, AARD and root mean square error (RMSE) are applied, and the latter is defined as:

$$RMSE = \sqrt{\frac{1}{m}\sum_{j=1}^{m}(y_{p,j} - y_{p,j}^*)^2} \quad (15)$$

The RMSE as an evaluation metric aims to mitigate potential biases that may arise from the AARD metric, especially when the response values approach zero.

Figure 9 shows the RMSE and AARD of the cold-flow (mixing) results predicted by the parametric ROM with different ML methods at two test points. Different ML methods exhibit good performance with AARD less than 10%, with the exception of SVR. The ROM with kriging presents an overall superior performance in predicting almost all physical fields and across different test datasets with the AARD of $X_{O_2}$, $X_{H_2}$, $X_{CH_4}$, and velocity magnitude less than 1.2%, 3.0%, 4.0% and 2.1%, respectively. The ROMs with MPR and DNN have similar RMSE and AARD for most of the physical variables. A common impression is that the nonlinear relationship between input and output can be better captured by DNN than other simple ML methods, such as KNN and MPR. ROM with KNN, however, predicts oxygen composition better than with DNN in both test cases. The underlying reasons may be two-fold: (1) DNN usually requires a gigantic database to be well performed, which cannot be guaranteed since many engineering problems are specific and associated data availability is limited; (2)

The ML methods like KNN rely closely on the relative position of test points in the sampling space, and the accuracy of the KNN method could downgrade in case the neighboring points in the sampling space are distant away from the test point of interest.

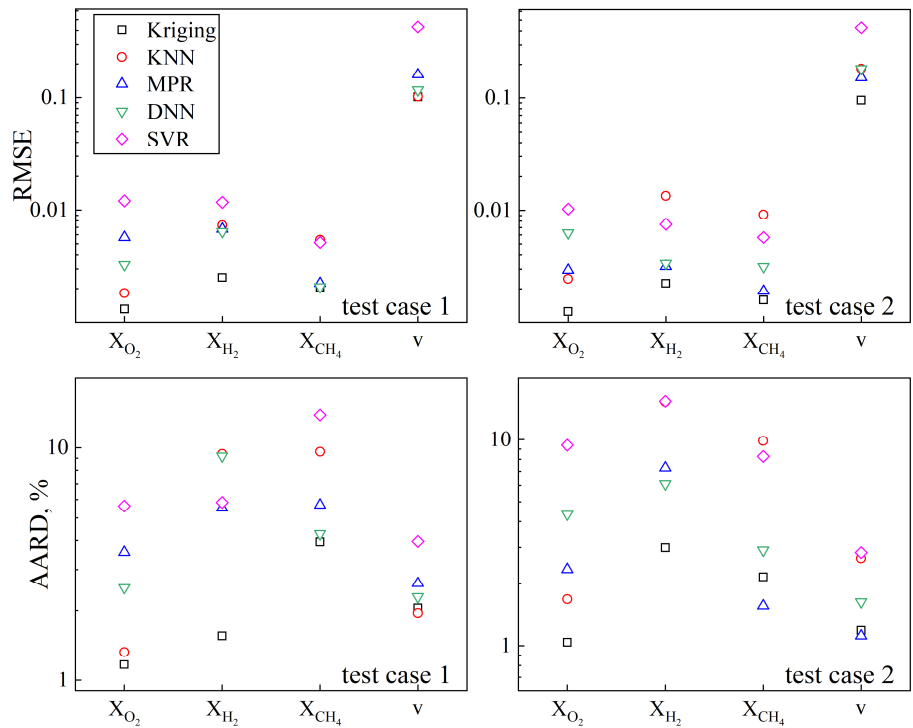

Fig. 9 RMSE (upper) and AARD (lower) of physical fields predicted by parametric ROM with different ML methods

Figure 10 presents the spatial distributions of hydrogen mole fraction, with discrete contour lines predicted by the parametric ROM with ML methods of kriging, MPR, KNN, and DNN in test case 1. For comparison, the upper half represents the results of various ML methods, while the lower half corresponds to the RANS counterpart, separated by a dashed line. The results demonstrate that the ROMs with kriging and DNN can effectively capture the spatial distribution of hydrogen mole fraction and the length of the potential core, compared to those with KNN and MPR. The performance of kriging is slightly better than that of DNN. This is evident from the degree of alignment of the contour lines between the RANS field and the ROM-predicted field.

Figure 11 shows the predicted distribution of oxygen mole fraction by the parametric ROM with various ML methods, as compared to the RANS results in test case 1. The oxygen jet leans towards the centerline owing to the entrainment of the central fuel jet. Kriging exhibits superior prediction accuracy to other ML methods, while KNN presents better performance than DNN in the vicinity of the contour line of 0.2. This observation is consistent with the comparison of RMSE and AARD in Fig. 9. Similar conclusion is found in the analysis of test case 2 (not shown here).

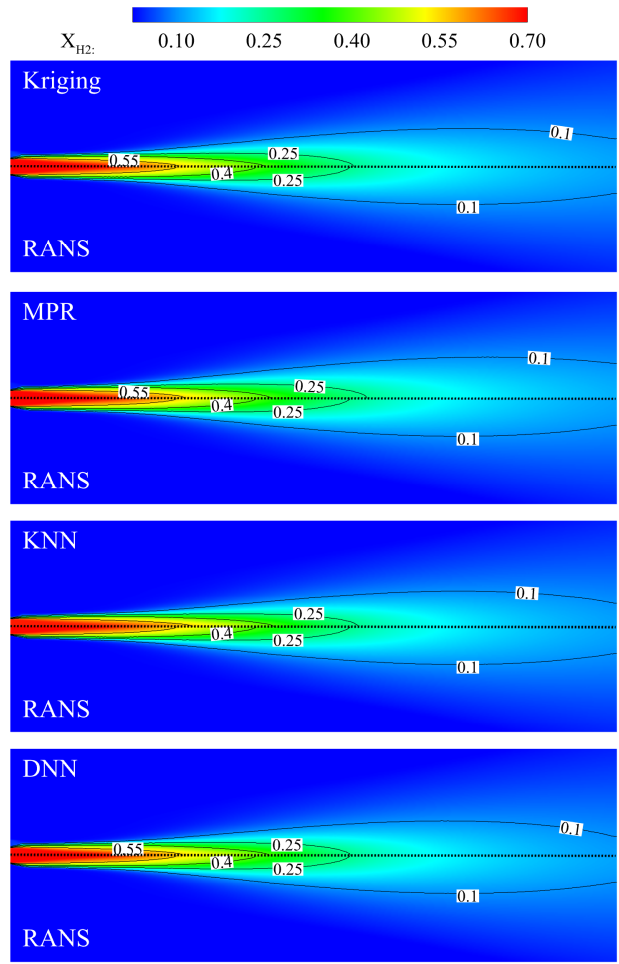

Fig. 10 Comparison of hydrogen mole fraction field: RANS (lower) and ROM with four ML methods (upper)

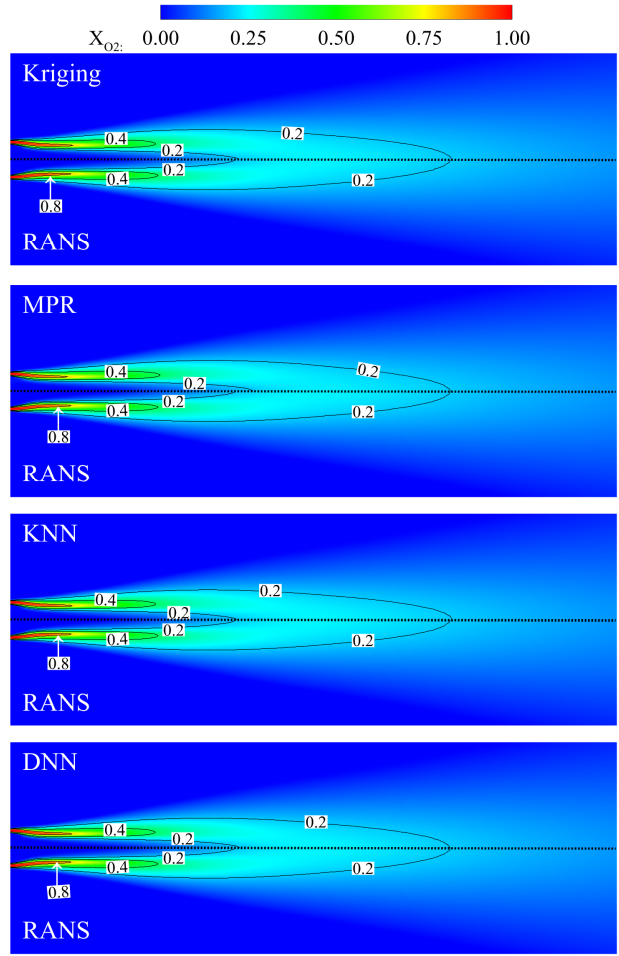

Fig. 11 Comparison of oxygen mole fraction field: RANS (lower) and ROM with four ML methods (upper)

To examine the performance of various ML methods more quantitatively, Figure 12 illustrates the radial distribution of hydrogen mole fraction predicted by different ML methods at various axial locations. The radial species gradient decreases along the streamwise direction, implying improved mixing when convecting downstream. All ML methods display close agreement with RANS at the nozzle near field ($x$=0.10 m), but the difference becomes larger in the downstream region except for kriging. Both KNN and DNN underestimate hydrogen concentration while MPR overestimates it. The POD-based ROM with kriging (red dashed line) reveals excellent performance throughout the flowfield, with a maximum relative error of only 2.08%.

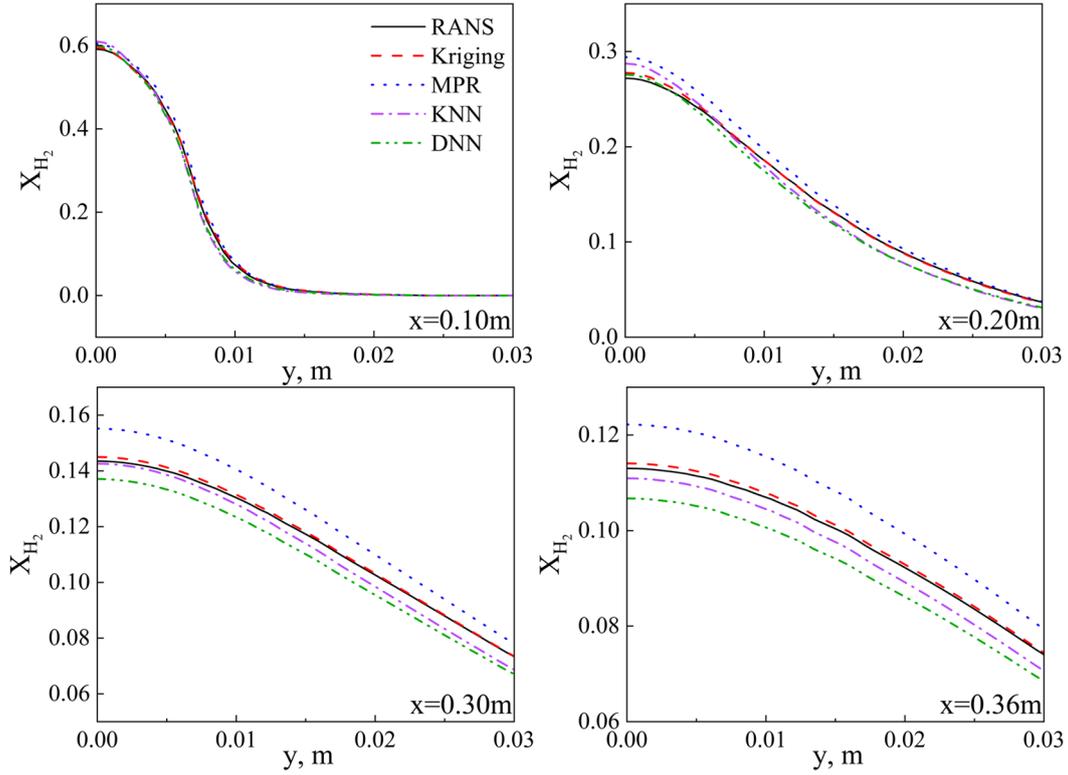

Fig. 12 Comparison of hydrogen mole fraction between ROM with various ML methods and RANS along the radial direction

To establish broad confidence, the ROM with kriging is assessed for all physical fields of interest in test case 2. Figure 13 displays the predicted results compared to the RANS counterpart. The flow structures and the spatial distribution of physical fields are well captured. The associated RMSE and AARD have been shown in Fig. 9. Note that RMSE by kriging is less than that by MPR for the prediction of methane composition and velocity magnitude, but AARD shows the opposite trend. This may be caused by the locally small magnitude of the denominator when calculating AARD, which could amplify the relative error.

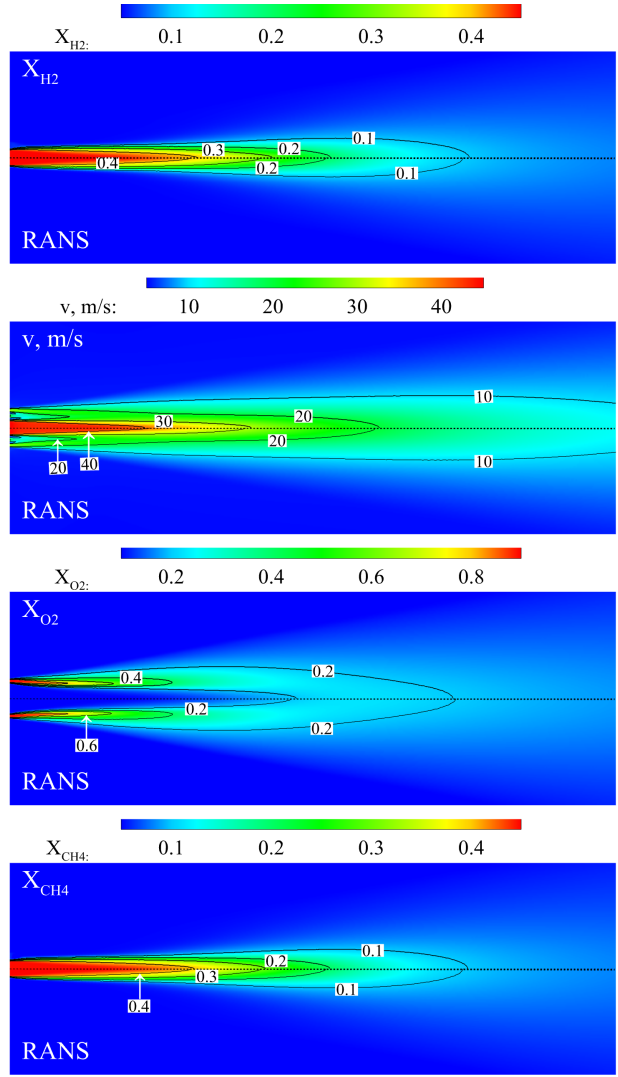

Fig. 13 Spatial distributions of $X_{H_2}$, velocity magnitude, $X_{O_2}$ and $X_{CH_4}$ from RANS-based simulation (lower) and POD-based ROM with kriging

### 4.3. ROM of combustion case

In this subsection, the parametric ROM with different ML methods is further employed for combustion cases. Figure 14 presents the predicted distribution of velocity magnitude compared to the RANS result in test case 2. Kriging and KNN can capture the spatial patterns of the velocity magnitude relatively well, while MPR and DNN fail to sufficiently represent the spatial field, as manifested by discrete contour lines in the figure.

The accurate prediction of intermediate species poses significant challenges due to their relatively small magnitude. In our study, we tend to create separate ROMs to emulate the spatial field of radicals, such as $CH_2O$ and OH. The latter (OH) represents the local reactive zone with significant heat release in the flame field [46]. For example, Figure 15 compares the predicted results of $X_{OH}$ by different ML methods in test case 1. Consistent with the results of the velocity magnitude field, both kriging and KNN predict the reactive zone well. The axial distance of flame initiation to the nozzle exit (related to lift-off distance) is accurately captured. MPR and DNN, however, underestimate substantially the lift-off distance and the maximum concentration of OH radical.

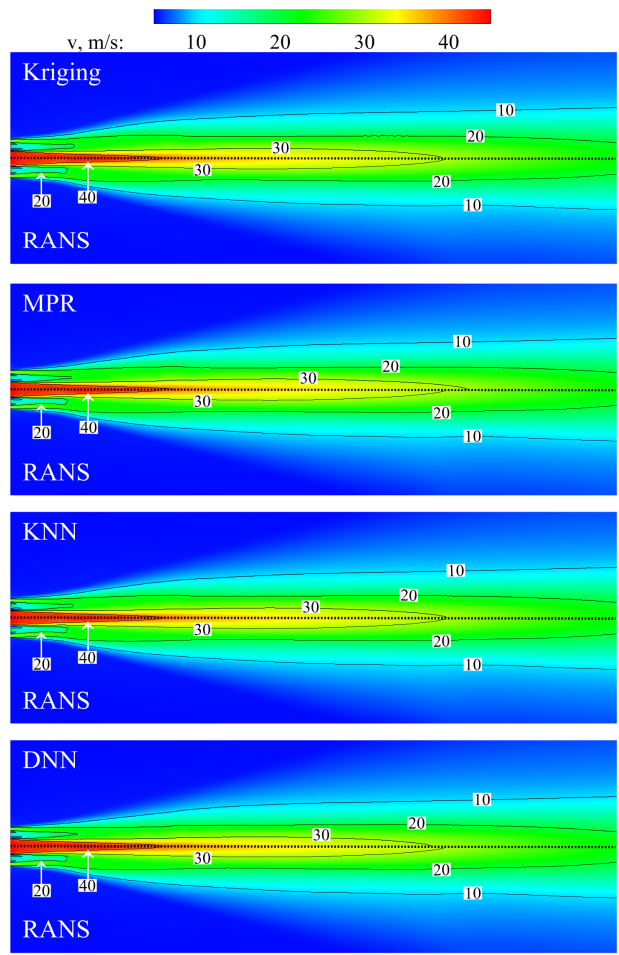

Fig. 14 Comparison of velocity magnitude field: RANS-based simulation (lower) and POD-based ROM with four ML methods (upper)

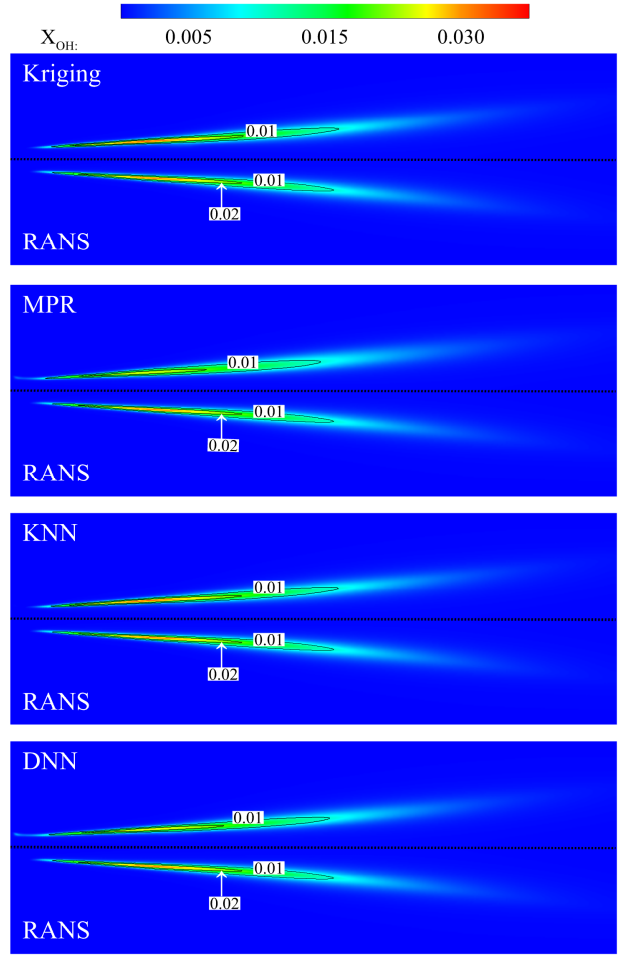

Fig. 15 Comparison of $X_{OH}$ field: RANS (lower) and ROM with four ML methods (upper)

Figure 16 provides AARDs of all physical fields of concern using ROM with different ML methods. It is seen that AARDs in combustion cases are higher than those in mixing cases, implying the challenge of predicting reacting flows using ROM caused by inherent nonlinearity embedded in the problem. The kriging method exhibits an AARD lower than 10% for all physical variables in both test cases. For some variables, KNN shows slightly lower AARD than kriging, while for CO2 in test case 1 KNN presents the least prediction accuracy among all ML methods. SVR and MPR have overall the poorest performance in two test cases, and DNN does not present its unique advantage claimed in the literature owing to the argument discussed in the previous section. In summary, the ROM with kriging consistently exhibits excellent performance across almost all physical fields in test cases with strong robustness and good reliability.

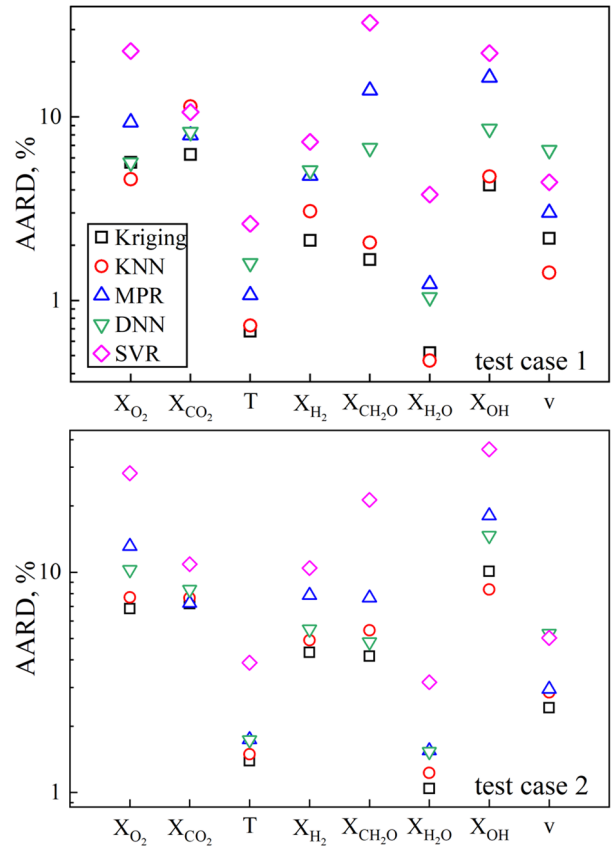

Fig. 16 AARD (%) of physical fields in different ML models

To further illustrate the performance of kriging, the reacting flowfields predicted by ROM with kriging in test case 1 are examined. Figure 17 shows the spatial distributions of temperature, $X_{CH_2O}$, $X_{CO_2}$ and $X_{H_2}$. From a physical perspective, $CH_2O$ radical serves as a significant precursor in controlling the initiation of the reaction, peaking prior to ignition. The associated AARDs have been shown in Fig. 16. The AARD of $X_{CO_2}$ is relatively high at 6.27%, while the performance for the spatial distribution of hydrogen mole fraction with an AARD of 2.13%. In general, the ROM with kriging provides accurate predictions for the physical fields of interest. This observation aligns with the conclusion obtained from the mixing problem, thereby confirming the significant advantage of the kriging method for highly nonlinear problems involving multi-scales and chemical reactions.

To have a more quantitative evaluation, the distribution of temperature in test case 2 and $X_{CH_2O}$ in test case 1 along the radial direction are shown in Figs. 18 and 19, respectively. Overall, the ROM with kriging demonstrates exceptional performance, and the predicted temperature and $X_{CH_2O}$ curves (red dashed lines) closely attach to the RANS (black solid lines) counterparts with a maximum relative error of less than 2.2% and 1.94%, respectively. This further substantiates the preeminence of the kriging method.

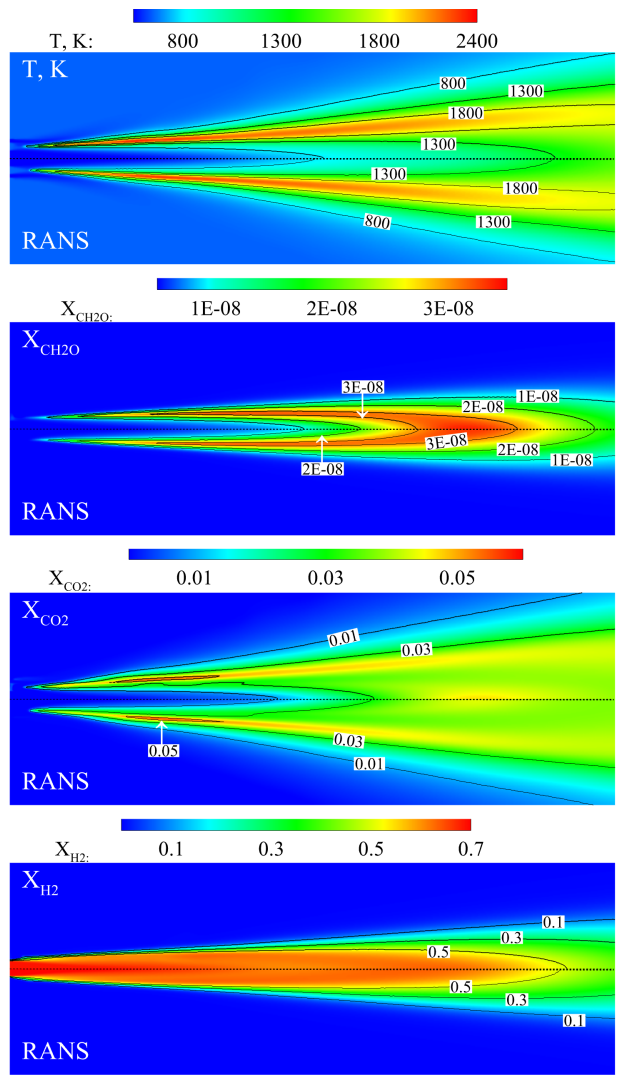

Fig. 17 Comparison of temperature, $X_{CH_2O}$, $X_{CO_2}$ and $X_{H_2}$ field: RANS (lower) and POD-based ROM with Kriging (upper)

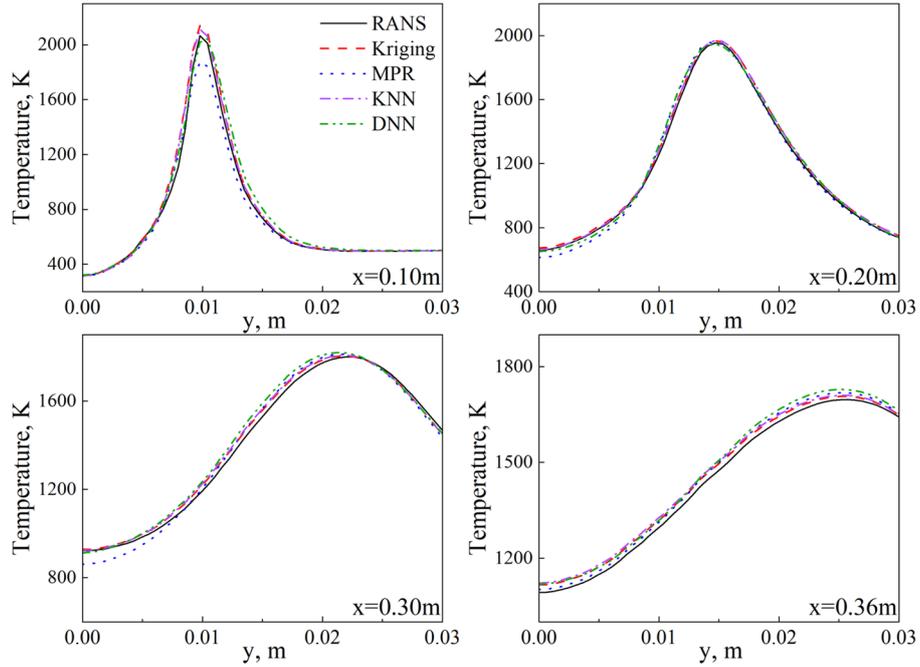

Fig. 18 Comparison of temperature distribution along the radial direction

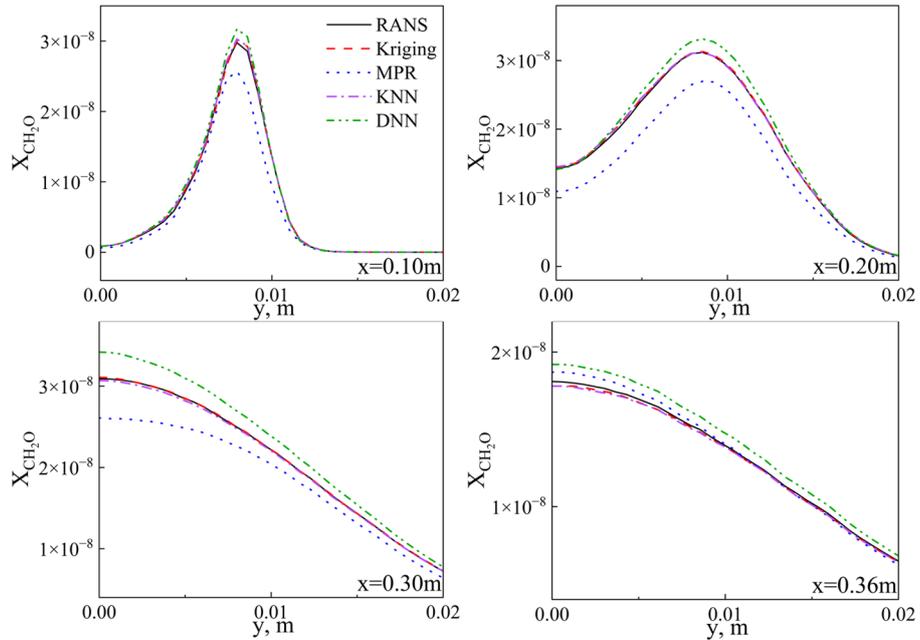

Fig. 19 Comparison of $X_{CH_2O}$ distribution along the radial direction

### 4.4. Computation performance

Computation efficiency is a crucial factor for engineering design and optimization. One of the motivations for developing parametric ROM is to substantially reduce the computation load of a physical problem to facilitate decision making during combustor design process. For the present problem, the time required to build the parametric ROM is about 30 seconds on 1 CPU (2.1GHz Intel Xeon Gold 5218R), which primarily is taken up by the eigendecomposition process. The time for POD coefficient training using ML methods and ROM prediction is minimal. Figure 20 shows the comparison of training and prediction times for the ROM with different ML methods. The training time refers to the time required to train the ML models using design input parameters and their corresponding POD coefficients. The prediction time refers to the time required to predict the POD coefficients at new parameters and to emulate the spatial flow field. Both times are obtained by iterating the process 5000 times and tested on 1 CPU. Among the different ML methods, kriging demonstrates the fastest training speed, and DNN is the lowest. In the prediction stage, KNN requires more prediction time due to the extraction procedures for finding the nearest $k$ sample points. The computational time for the RANS-based simulations in the current study ranges from 6 to 12 hours using 48 CPUs for different design settings. Therefore, in the case of having a pre-trained database at hand, the prediction of spatial fields by a POD-based parametric ROM can achieve time savings of up to eight orders of magnitude, compared to RANS-based simulations.

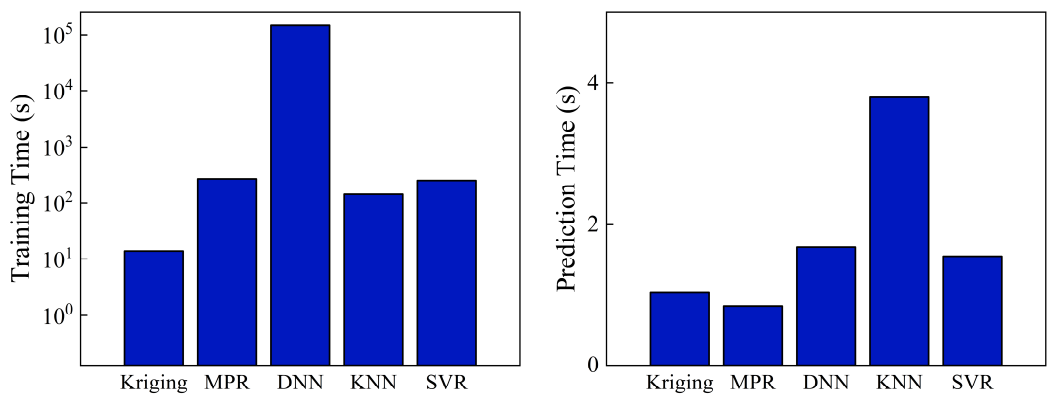

Fig. 20 Comparison of training and prediction time for different ML methods (5000 iterations)

## 5. Conclusion

In this study, parametric ROMs using various machine learning (ML) methods are proposed for emulation of spatial distributions of physical fields for multi-species mixing and combustion problems. The parametric ROMs, incorporating experimental design, assimilation of high-dimensional numerical data, proper-orthogonal-decomposition (POD)-based model reduction, and various ML techniques, can effectively serve as a surrogate of the full-order CFD flow solver in the prescribed parameter space. The performance of parametric ROMs is evaluated through the emulation of multi-species mixing and combustion of fuel blend and oxygen issuing from a triple-coaxial nozzle in a steam-diluted environment, with fuel blending ratio (hydrogen/methane) and steam dilution ratio selected as design parameters. A total of 20 training design points are selected using the Latin hypercube design approach, and two validation cases are included. Data is assimilated through Reynolds-averaged Navier-Stokes (RANS)-based simulations of training and validation cases.

The results show that the POD-based parametric ROM with kriging provides superior performance in predicting almost all physical fields, including temperature, velocity magnitude, and combustion products, at different validation datasets, compared to other ML techniques. The accuracy of ROM with KNN also performs well in the region near the sampling design points, whereas the SVR and MPR have relatively poor performance. The accuracy of ROM with DNN is not encouraging, as DNN training typically necessitates a large amount of data, which cannot be guaranteed due to the specificity of many engineering problems and associated data are spare and expensive. In regard to computation efficiency, the parametric ROM can significantly reduce the computing time of predicting a spatial field at a new design point by eight orders of magnitude, as compared to RANS-based simulation. The present framework provides a novel pathway for fast emulation of mixing and combustion problems that are conventionally based on numerical simulations, and can also be applied to a wide range of engineering applications.

## Acknowledgements

This work was supported by the Science Center for Gas Turbine Project (P2022-B-II-020-001, P2021-A-I-003-002) and National Science and Technology Major Project (Y2019-I-0022-0021).